%% file: AGCOM_arxiv_version.tex
\documentclass[letterpaper]{IEEEtran}
\IEEEoverridecommandlockouts
\makeatletter
\def\markboth#1#2{\def\leftmark{\@IEEEcompsoconly{\sffamily}\MakeUppercase{\protect#1}}%
\def\rightmark{\@IEEEcompsoconly{\sffamily}\MakeUppercase{\protect#2}}}
\makeatother

\usepackage[acronyms,nonumberlist,nopostdot,nomain,nogroupskip]{glossaries}
\input{acronyms.tex}

\usepackage{kantlipsum}
\usepackage{setspace}

\usepackage[english]{babel}
\usepackage{xcolor}
\usepackage{lipsum}
\usepackage{caption}
\usepackage{cite}
\usepackage[pdftex]{graphicx}
\usepackage{subcaption}
\usepackage{amsmath}
\usepackage{booktabs}
\usepackage{colortbl}
\usepackage{amsfonts}
\usepackage{amssymb}
\usepackage{amsthm}
\usepackage{array}
\usepackage{verbatim}
\usepackage{listings}
\usepackage{algorithm}
\usepackage{algpseudocode}
\usepackage{url}
\usepackage{enumerate}
\usepackage{multirow}
\usepackage{xfrac}

\definecolor{LightBlue}{rgb}{0.5,0.5,1}
\definecolor{LightRed}{rgb}{1,0.5,0.5}
\definecolor{LightYellow}{rgb}{1,0.85,0}

\usepackage{epsfig}
\usepackage{epstopdf}
\usepackage{multicol}
\usepackage[font=footnotesize]{caption}

\usepackage{algorithm}

\makeatletter
\def\BState{\State\hskip-\ALG@thistlm}
\makeatother

\renewcommand{\arraystretch}{2}

\def\3G{\mathsf{3GPP}}

\title{A Spectrum Sharing Solution for the Efficient Use of mmWave Bands in 5G Cellular Scenarios}
\author{{{\textbf{Mattia Rebato\thanks{The authors would like to thank Dr. Federico Boccardi for useful discussions about this work. The work of M. Zorzi has been partially supported by NYU Wireless.}} and \textbf{Michele Zorzi}} }\\ \normalsize Department of Information Engineering,
University of Padova, 35131 Padova, Italy \\ \small{$\{$\texttt{rebatoma, zorzi}$\}$\texttt{@dei.unipd.it}} \vspace{-0.5cm}}

\makeglossaries
\begin{document}
\maketitle

\thispagestyle{empty}

\begin{abstract}

Regulators all around the world have started identifying the portions of the spectrum that will be used for the next generation of cellular networks.
A band in the mmWave spectrum will be exploited to increase the available capacity.
In response to the very high expected traffic demand, a sharing mechanism may make it possible to use the spectrum more efficiently. 
In this work, moving within the European and Italian regulatory conditions, we propose the use of \gls{lsa} to coordinate sharing among cellular operators.
Additionally, we show some preliminary results on our research activities which are focused on a dynamic spectrum sharing approach applied in simulated 5G cellular scenarios. 
\end{abstract}

\begin{IEEEkeywords}
Millimeter-wave, 5G, spectrum sharing, spectrum optimization, QoS, LSA, spectrum broker.
\end{IEEEkeywords}
\begin{picture}(0,0)(-16,-295)
\centering
\put(0,0){
\put(-28,10){ M. Rebato and M. Zorzi, ``A Spectrum Sharing Solution for the Efficient Use of mmWave Bands in 5G Cellular Scenarios",}
\put(-13,-0){in IEEE International Symposium on Dynamic Spectrum Access Networks (IEEE DySPAN), Seoul, Korea, Oct 2018.}}
\end{picture}

\glsresetall
\glsunset{nr}

\vspace{-0.5cm}
\section{Introduction}
\label{introduction}

The next generation of cellular networks will need to cope with a very high mobile traffic demand, due to the expected increase in the number of connected devices and of the traffic they produce~\cite{cisco2017}.
As an enabler for these capacity-intensive applications, the mmWave band (approximately between 10 and 300 GHz) has been identified as a promising candidate for communication, thanks to the availability of wide portions of free spectrum~\cite{rangan2017potentials}.
Therefore, the fifth generation of cellular networks (5G), which is currently being standardized by the \gls{3gpp}, will introduce carrier frequencies in the mmWave bands.

In the meantime, in addition to the \gls{3gpp} specification, spectrum regulators are providing indications on the mmWave bands they plan to release, and on the authorization mechanisms they plan to use in these frequencies. 
In the European spectrum specifications~\cite{rspg18}, each \emph{Member State} should indicatively consider the range of frequencies between 24.25 and 27.5~GHz in a way to uniformly spread the use of 5G frequencies throughout Europe. 
Furthermore, each \emph{Member State} is required to be flexible in the mix of authorization approaches to use.
Alternative authorization approaches may include general authorization regimes (license exemption), exclusive license, licensed shared use between different \glspl{ue}, geographical sharing (comprising sub-national, regional and site-specific licensing, including at the local level directly to businesses), or more dynamic approaches to spectrum sharing in time and space, possibly using geolocation databases~\cite{rspg18}.

Another important aspect, which should be properly settled, regards the coexistence of 5G systems around the 26~GHz band (or equivalent mmWave bands) with other services such as wireless fixed links and also \glspl{fss}~\cite{guidolin}.
Depending on the location of the fixed links, the demand for 5G small cells, and the extent to which interference can be mitigated using new technologies, it may be possible to deploy 5G small cells within the same frequency range as some of these existing fixed links.

\begin{figure}[t!]
\centering
\includegraphics[width=0.6\columnwidth]{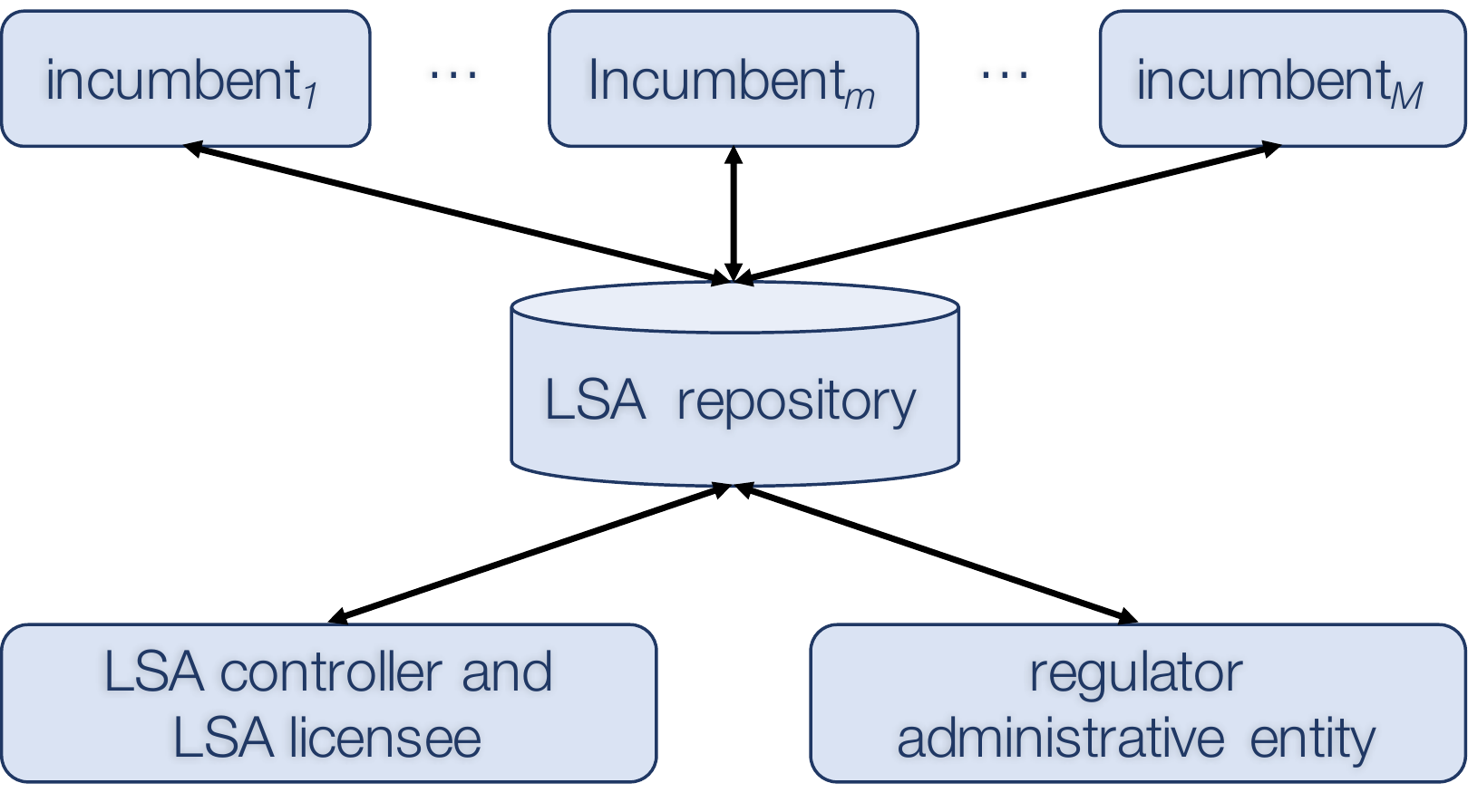}
\caption{The \gls{lsa} architecture reference model~\cite{massaro}.}
\vspace{-0.5cm}
\label{lsa_block_scheme}
\end{figure}

Even if the mmWave spectrum is large, in order to fit multiple operators in the available band, high 5G performance will only be possible with intelligent spectrum management mechanisms.
For this reason, there arises a question from the regulators on how to efficiently use the available spectrum.
According to regulatory rules which impose the full use of the band, and in order to address the question of how to properly use the available spectrum, in this work we propose a spectrum sharing solution for the mmWave band in 5G cellular scenarios.
This requires an adaptive technology which can control and coordinate the sharing between operators.
This technology must operate with a known language so that it can be used by all the entities in the network.
To be precise, in this work we suggest the use of either a \gls{lsa} approach or similarly the use of a third-party spectrum broker, both with the role to control and dynamically coordinate the sharing of licensed spectrum and efficiently use the resources according to operators needs.
For example, if in a particular spatial and time instance the traffic grows faster and traditional exclusive spectrum is not enough, \gls{lsa} allows for sharing while meeting the requirements of mobile operators and incumbents for predictable conditions of spectrum use and \gls{qos}.
The accessible band of any operator can be dynamically adjusted according to the needs of each network in the environment at any particular instant. 
Efficient spectrum sharing is necessary to provide fairness in the allocation as well as service satisfaction across multiple \glspl{ue} while maximizing the spectral efficiency and the utilization of the total available bandwidth.
In order to use the available spectrum in an efficient way, there should not be any unused resources.
\gls{lsa} may be a proper approach to avoid this issue, as it facilitates access to additional licensees in bands which are already in use by one or more incumbents\footnote{For example, such an incumbent may be a telephone service provider (i.e., an operator) that owns the license for the band.}.
We note that this should not be confused with the \gls{laa} approach, a technology which involves the sharing and aggregation of different bands (e.g., unlicensed spectrum) or also different Radio Access Technologies.
Instead, \gls{lsa} is a concept which permits to dynamically share the band, whenever and wherever it is unused by the incumbent users, i.e., \glspl{pu}.
The shared use of the spectrum is only allowed on the basis of an individual authorization (i.e., operators holding an \gls{lsa} license or registered to use the band).
As shown in Fig.~\ref{lsa_block_scheme}, each operator (i.e., incumbent) can request an \gls{lsa} license through the \gls{lsa} repository. 
Then, the \gls{lsa} controller, in agreement with the administrative entity, has the task to address all the license requests. 
\gls{lsa} is defined within the framework of the European Union as: \emph{``a regulatory approach aiming to facilitate the introduction of radiocommunication systems operated by a limited number of licensees under an individual licensing regime in a frequency band already assigned or expected to be assigned to one or more incumbent users"}~\cite{rspg18}.
Under this approach, the additional users are authorized to use the spectrum (or part of the spectrum) in accordance with sharing rules included in their rights of use of the spectrum, thereby allowing all the authorized users, including incumbents, to provide a certain~\gls{qos}~\cite{rspg18}.

\vspace{-0.4cm}
\subsection{Related Works}
Works~\cite{future_spectrum,Bhattarai} and~\cite{survey} discuss possible sharing approaches that can be used for future mobile scenarios.
However, in all these survey works simulation results are missing and a detailed mmWave environment was not considered.
Some simulations and numerical evaluations were presented in~\cite{rebato16,boccardi16} and~\cite{rebato_tccn}.
In both~\cite{rebato16} and~\cite{boccardi16} it is shown that spectrum sharing at mmWave has the potential for a more efficient spectrum use than a traditional exclusive spectrum allocation to a single operator.
Work~\cite{rebato_tccn} introduces a hybrid spectrum access scheme for mmWave networks, where data packets are scheduled through two mmWave carriers with different characteristics.
In particular, the authors combined a lower mmWave band with exclusive access and a higher mmWave band where spectrum is pooled between multiple operators.
As a result, the investigation shows that this approach provides advantages for the average \gls{ue} with respect to traditional fully licensed or fully pooled spectrum access schemes. 
The approach in~\cite{rebato_tccn} is dynamic, but cannot change during transmission, which means that the spectrum allocation is done only once, e.g., before starting the transmission.
Differently, work in~\cite{dynamic_hspsh}~reports a comparison between fixed and dynamic spectrum sharing and shows that dynamic spectrum sharing can benefit from spectrum handoff to enhance the rate performance by switching from the unavailable channels to the available ones, thereby maximizing the utilization of the total available bandwidth.
Generally, dynamic spectrum sharing can benefit from spectrum handoff.
On the other hand, static spectrum sharing can avoid the impacts of spectrum handoff delay by allowing \glspl{su} to back off and wait if any \gls{pu} is using the same channel.
In~\cite{merwaday}, the authors proposed a spectrum market mechanism where sharing is promoted explicitly by the government which regulated the use of the spectrum.
The frequency regulator offers subsidy support to the wireless operators and requires a performance metric to be reported.
Therefore, the spectrum is better exploited and all the entities benefit from this approach.
Finally, work~\cite{luo} proposed the use of a geolocation database together with a spectrum broker to control the time and spatial allocation~of~the~band. 

None of the above related works have discussed the possibility to apply an \gls{lsa} approach or a third-party spectrum broker as a controller which helps to improve the spectral efficiency of the networks in mmWave bands.
For this reason, in this work we are suggesting the use of such approaches for mmWave bands in a way to improve the spectral efficiency and the \gls{qos} that operators can ensure to the customers.    

\vspace{-0.35cm}
\section{Spectrum Use}
\label{system_spectrum_model}

\subsection{Sharing Mechanisms}
\label{sharing_mechanisms}

Before entering into the details about the Italian spectrum specifications\footnote{We are considering the Italian spectrum regulator because it was one of the first agencies setting conditions on the use of mmWave bands for 5G~\cite{agcom_doc}.}, we discuss in this first part the two possible mechanisms that can be used to dynamically allocate spectrum.
On one hand, we can consider an \gls{lsa} approach where the administrative entity has direct control of the licensed band and also the temporal \gls{lsa} licenses which are distributed to the other operators, i.e., the secondary users (\glspl{su}).
On the other hand, a spectrum broker can also be considered which is not directly controlled by an administrative entity, but rather by a third-party company.
In this second approach, only an agreement between operators is required, so that more complex business models become possible.
It is clear that both these allocation procedures depend on the spatial region considered.
Therefore, a geolocation database will be needed to store information about which portion of the band is shared and in which region.

Furthermore, an additional economic study is required, in order to understand if operators have business advantages to dynamically share the spectrum.
In this preliminary analysis, we are not focusing on any economic aspects.
Therefore, we reserve as a future work the study of how to optimize the use of spectrum considering also a cost model for sharing the~band. 

In the following section, we will analyze in detail the Italian spectrum specifications for the 5G mmWave band.
However, similar procedure and considerations can be applied to other bands according to their corresponding regulatory specifications. 

\vspace{-0.4cm}
\subsection{Italian regulatory conditions}
\label{italian_spectrum}
\begin{figure}[t!]
\centering
\includegraphics[width=0.8\columnwidth]{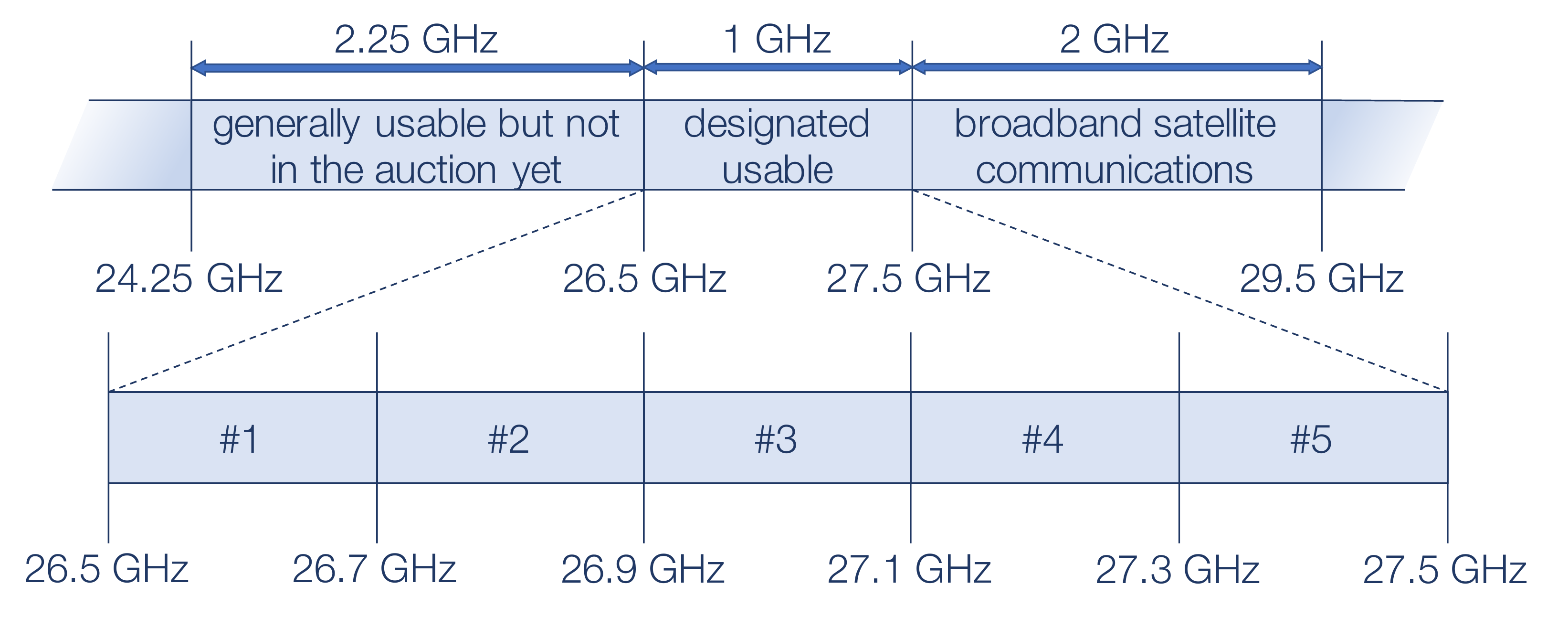}
\caption{View of the spectrum around 27~GHz. The top part shows the designated use for each band, while in the bottom we show the five chunks of 200~MHz each, which will be auctioned in the Italian territory as specified in~\cite{agcom_doc}.}
\vspace{-0.5cm}
\label{spectrum_scheme}
\end{figure}

AGCOM, the Italian spectrum regulator, provides a plan in~\cite{agcom_doc}, where 1~GHz of bandwidth at 26~GHz is designated for 5G applications and allocated for a future auction\footnote{Even though the document refers to this portion of spectrum as the \emph{26~GHz band}, the precise portion of spectrum is between 26.5 and 27.5~GHz.}.
Furthermore, the entire bandwidth is slotted in five chunks with fixed sizes of 200~MHz each. 
Each operator can buy the license for at most two chunks and, if a band is not used, other operators or services in the area can use the portion of unused spectrum.
This last rule is fundamental in order to efficiently use all the spectrum and avoid waste of resources.
This restriction is valid in relation with the area considered, and therefore the use of the spectrum may vary in different regions.  
For this reason, an approach such as \gls{lsa} combined with geo-location databases appears to be a proper solution to address the use of resources.
A display of the portion of the band in question can be seen in Fig.~\ref{spectrum_scheme}.
The figure exhibits also the current use of the frequencies below 26.5~GHz and above 27.5~GHz (e.g., the broadband satellite communications designated for frequencies between 27.5 and 29.5~GHz).
The detailed values here reported are taken from~\cite{agcom_doc} and are valid for the Italian territory.
Even if the concept proposed in this work is studied focusing on those values, it can be identically considered in others portion of the spectrum for the other European Member States and likewise other countries outside Europe.

As a possible result of the auction, five operators can buy a chunk of band each or, differently, four operators can buy the license of a chunk each, while the last unsold chunk can be shared among all the operators. 
Another possible outcome of the auction can see a single operator owning two chunks (i.e., 400~MHz of band) and the other three owning only a single chunk each.
These are all possible outcomes of the auction.
In the next part of the paper we will focus on the evaluation of just the first case, which is the most appropriate choice.
Thus, we are considering five operators assigned a chunk of 200~MHz each.

Following these preliminary assumptions, we examine here a scenario with $M = 5$ operators indexed by $m \in \{1,\ldots,M\}$.  
Each operator owns distinct \glspl{gnb} with no particular infrastructure sharing between operators, and the only iterations take place through the \gls{lsa} controller.
If $W_{tot}= 1$~GHz is the total system bandwidth available in the portion of spectrum accessible for auction (i.e., band around 26~GHz), initially we assume that each of the five operators is auctioned a $W_m = 200$~MHz chunk of licensed bandwidth, in which the $m$-th operator is the \gls{pu} and has priority access.
Therefore, bands assigned to different operators are disjoint.
In this exploratory study, the effective allocation is dynamically adjusted according to the instantaneous traffic. In particular, an operator whose traffic is below a certain threshold must release part of its resources to other operators on a secondary basis. 

As previously mentioned, in this paper we propose and evaluate an approach where the spectrum is dynamically allocated.
The use of the dynamic access scheme is motivated by the fact that in some cases an operator may need more resources with respect to the other operators in a particular area.
Thus, a dynamic allocation of the resources can help to improve the \gls{qos} experience of the \glspl{ue} in a particular time and space instance.
The dynamic approach offers also better balancing of the available resources with respect to the baseline case, resulting in higher fairness.
By adding the dynamic component, the average throughput and spectral efficiency are further improved thanks to the dynamic allocation of resources among operators. 
We note that this approach is reminiscent of the \gls{lsa} framework, in fact, the dynamic component of the approach is achieved thanks to the \gls{lsa} procedure, which permits to share the unused \gls{pu} resources with others operators (i.e., \glspl{su}) as a way to better use the entire bands.
Similarly, as previously mentioned, this dynamic sharing can be managed by a third-party broker which has the same distributed role of controlling the use of resources as in the \gls{lsa} mechanism.
In order to reproduce this behavior, we model a scenario where the allocated band for each operator is proportional to its traffic.

\vspace{-0.35cm}
\section{System Model and Evaluation Methodology}
\label{simulation_methodology}
In order to provide a proof-of-concept evaluation of the proposed dynamic hybrid spectrum access approach under realistic scenarios for mmWave cellular systems, we study it through a simulation methodology, where detailed models are used for all important effects and variables (including in particular channel characteristics and association policies), as described below.
To do this, we simulate a dense area where multiple operators are co-located, and compare the performance of a baseline configuration with one based on dynamic sharing.

\paragraph*{Deployment Model}
For each operator $m \in \{1,\ldots,M\}$, the positions of the \glspl{ue} and of the \glspl{gnb} are modeled according to two \glspl{ppp}, with densities $\lambda_{\text{\gls{ue}}}$ and $\lambda_{\text{\gls{gnb}}}$ in area $A$.
This corresponds to considering an unplanned deployment, where \glspl{gnb} are not optimally located.

\paragraph*{Channel Model}
The \gls{mimo} channel matrices are generated according to a statistical channel model
derived from a set of extensive measurement campaigns in New York City~\cite{akdeniz14}.
We capture the metric of interest from the typical user located in the center of the area.
With this method, we remove the border effects by considering all the interfering terms, thereby correctly evaluating the statistics of interest for the typical user.

\paragraph*{Beamforming} 
We model the antennas as a \gls{upa} with $\lambda/2$ spacing at both the \gls{gnb} and the \gls{ue}.
Furthermore, we precisely model the antenna radiation pattern following the \gls{3gpp} specifications, as done in~\cite{rebato18}.
This permits to carefully characterize the steering beams, and therefore to have a precise knowledge of the amount of power irradiated by the antenna arrays in all directions, thus accurately computing the desired and interfering signals.
Among other simplifications, this model assumes perfect beam tracking and the ability to form an arbitrary BF vector.
Therefore, we can generate a beamforming vector for any possible angle and we also assume perfect alignment between the beams of each \gls{ue} and its serving \gls{gnb}.

\paragraph*{Rate and Scheduling Model}
For simplicity, in this initial study we assume that the channel gain is flat across time and frequency.  
We consider beamforming with single-stream transmissions (i.e., we do not consider spatial multiplexing) to any one \gls{ue}.
Thus, we define with term $G_{ij}$ the desired gain between \gls{gnb} $i$ and \gls{ue} $j$. 
Similarly, we consider the gain $G_{ijk}$ from an interfering \gls{gnb} $k$ from the same operator $m$.
In this case, the \gls{ue} will experience a time-varying interference as the interfering \gls{gnb}
directs its transmissions to the different \glspl{ue} it is serving\footnote{Detailed explanations of the channel, antenna and gain characterizations and calculations can be found in~\cite{akdeniz14,rebato_tccn} and~\cite{rebato18}.}.
The \gls{sinr} is then given by
\vspace{-0.2cm}
\begin{equation}
\gamma_{ij} = \frac{\frac{P}{\ell_{ij}}G_{ij}}
{\sum_{k \neq i}  \frac{P}{\ell_{kj}}G_{ijk} + W_m  N_0},
\label{equation_sinr}
\end{equation}
where $P$ is the total transmit power from the \gls{gnb}, $N_0$ is the thermal noise power spectral density and $\ell_{ij}$ is the path loss between \gls{gnb} $i$ and \gls{ue} $j$ and is computed considering \gls{los}, \gls{nlos} and outage states~\cite{akdeniz14}.
The summation in the denominator of~\eqref{equation_sinr} is over all \glspl{gnb} $k$ in the band, including \glspl{gnb} of the same operator.
Note that, within the cell, we assume that \glspl{ue} are scheduled on orthogonal resources (e.g., in time or frequency) and hence there is no intra-cell interference. 

Using the \gls{sinr} expression in~\eqref{equation_sinr} we approximate the throughput for the $j$-th user ($\eta_j$) as follows
\vspace{-0.1cm}
\begin{equation}
\eta_j=  \frac{W_m}{N_{i}^{(m)}} \log_2\left( 1 + \gamma_{ij} \right),
\label{th_equ}
\end{equation}
where the total available resources, which are identified by the band $W_m$, are split among all the $N_{i}^{(m)}$ users of operator $m$ associated to the $i$-th \gls{gnb}.
The ratio between the total bandwidth $W_m$ and the number of users $ N_{i}^{(m)}$ associated to the specific carrier provides the average amount of resources allocated to the $j$-th user over time.

We are also interested in the evaluation of the achievable performance in the case in which the transmission is performed with a coordination mechanism which permits to properly avoid inter-cell interference. 
In this last particular case, the interference can be neglected and the \gls{sinr} formula in~\eqref{equation_sinr} can be approximated with the \gls{snr} expression as follows
\vspace{-0.2cm}
\begin{equation}
\hat{\gamma}_{ij} = \frac{\frac{P}{\ell_{ij}}G_{ij}}
{W_m  N_0}.
\label{equation_snr}
\end{equation}

\paragraph*{Carrier Association}
In cell and carrier association, each \gls{ue} $j$ must be assigned a serving \gls{gnb} cell $i$.
To be precise, each \gls{ue} is associated with the \gls{gnb} that provides the smallest path loss among all the available \glspl{gnb} of the operator.
Multiple \glspl{ue} can be associated with a single \gls{gnb}, while the \gls{gnb} serves only a single \gls{ue} per unit time slot according to a uniformly random scheduler.

\begin{figure}[t!]
\centering
\includegraphics[width=.8\columnwidth]{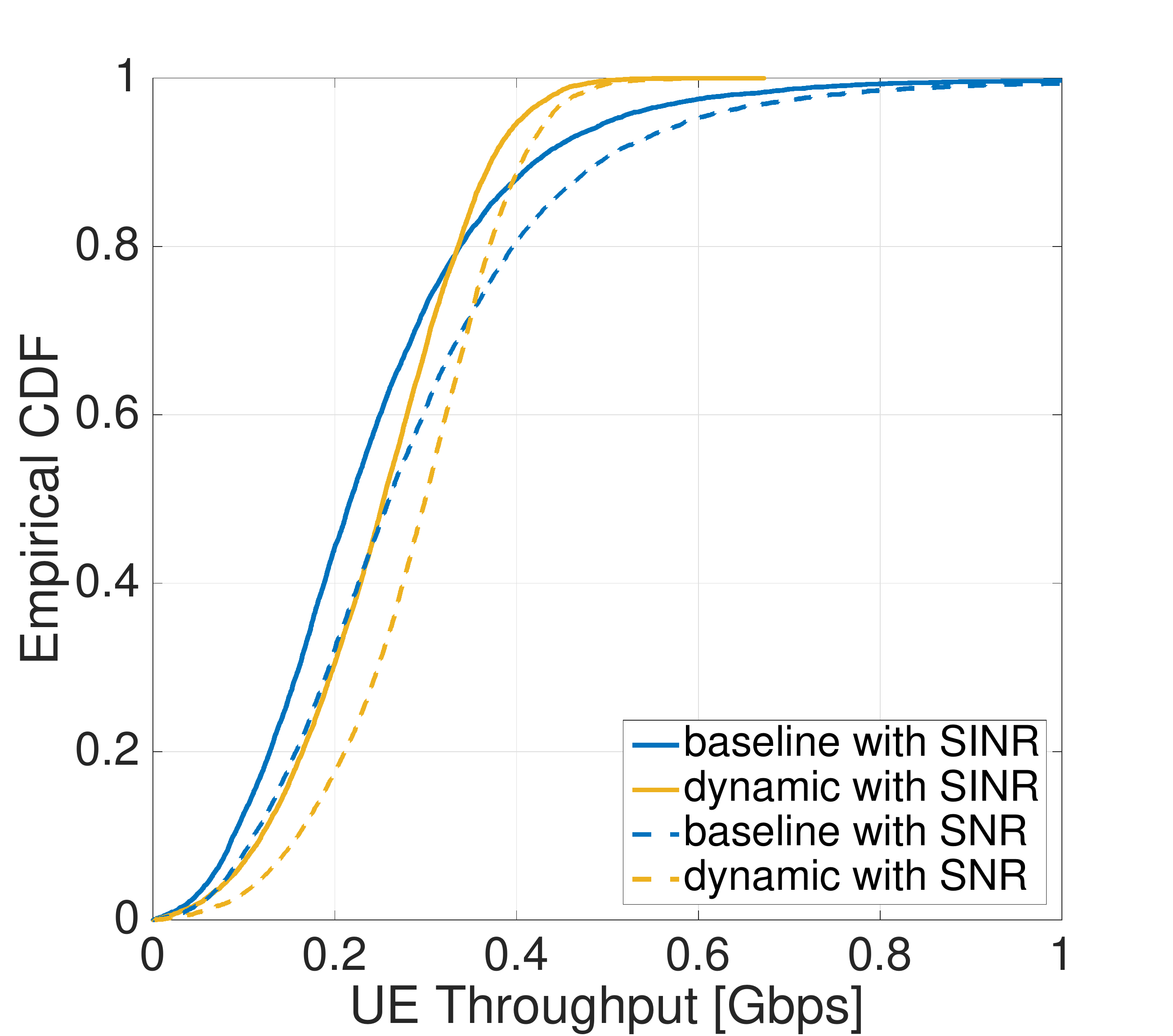}
\caption{Empirical \gls{cdf} of the throughput ($\eta$) per \gls{ue} in the two configurations evaluated. In particular, the dashed lines are used for throughput representations when interference is neglected, thus \gls{snr} is considered instead of \gls{sinr}. Figure obtained considering a \gls{gnb} density of 75~\gls{gnb}/km$^2$ for each operator.}
\label{fig_main_result}
\vspace{-0.5cm}
\end{figure}
\paragraph*{Evaluation Approach}
In order to evaluate the benefit of a dynamic sharing approach, we present a preliminary analysis of two different scenarios.
As the baseline case, we consider a scenario where all the operators are serving the associated \glspl{ue} using only the licensed chunk of band they own. 
As an alternative scenario, we consider a dynamic case where the resources are split among the operators according to the number of associated \glspl{ue} in the area. 
This simulation tool permits to understand the benefit and the achievable throughput of a system with a dynamic use of the resources. 
To be precise, defining as the typical \gls{gnb} the \gls{gnb} to which the typical \gls{ue} is associated, we compare the number of associated \glspl{ue} to the typical \gls{gnb} with respect to all the other \glspl{gnb} in the area.
We then split the total band $W_{tot}$ according to the need of each operator in the considered area. 
For instance, if the typical \gls{gnb} of operator $m$ has fewer associated \glspl{ue} with respect to the other \glspl{gnb} in the area, the allocated band $W_m$ used for the transmission of the typical \gls{gnb} will be smaller with respect to the licensed chunk operator $m$ has licensed.
Thus, the remaining part of the chunk will be shared and then allocated to other operators in the area with a larger number of associated \glspl{ue}.
Conversely, if the typical \gls{gnb} has more associated \glspl{ue} with respect to the other \glspl{gnb} in the area, the allocated band used for its transmission will be larger than the licensed chunk, therefore the use of some portions of bands licensed to other operators will be necessary.
We use the \emph{Jain} fairness measures to determine whether users are receiving a fair share of the system resources~\cite{jain}.
The fairness for a set of $n$ throughput values is computed as
\begin{equation}
\mathcal{J}(\eta_1,\eta_2,\dots,\eta_n) = \frac{(\sum_{i=1}^n \eta_i)^2}{n \sum_{i=1}^n \eta_i^2}.
\end{equation}
The metric $\mathcal{J}$ ranges from $\tfrac{1}{n}$, which represents the worst case, to 1 (best case), and it is maximum when all \glspl{ue} have the same throughput.

\vspace{-0.35cm}
\section{Numerical results}
\label{numerical_results}
We report in this section results obtained from the analysis we have done in order to understand the advantages that an efficient use of the spectrum can bring.  
As previously explained, we are comparing two different simulation scenarios: (\emph{i}) the baseline where the chunks of the band are equally and exclusively split among five operators; (\emph{ii}) a scenario where operator bands are shared and dynamically adjusted in relation to the needs of the operators in a particular region. 

We report in Fig.~\ref{fig_main_result} the empirical \gls{cdf} of the throughput for the typical \gls{ue} in all the compared configurations. 
As we can see, the procedure in which the band is dynamically adjusted according to the area results in better fairness among the \glspl{ue}.
In fact, the throughput is reduced for the best users (i.e., upper right part of the curve) while at the same time, passing from the blue curve to the yellow one, the throughput is improved for all the other users (e.g., worst and medium \glspl{ue}).
This means that the users in the dynamic procedure experience better fairness and the overall spectral efficiency increases. 
A similar behavior is achieved when the term $\hat{\gamma}_{ij}$ (i.e., \gls{snr})  is considered.
As expected, the throughput is bigger if an interference avoidance mechanism is adopted (as shown by the dashed curves).   
Furthermore, this last set of curves (i.e., when \gls{snr} is considered in place of the \gls{sinr}) allows us to understand the upper bound that can be reached in the case in which all the interference conditions can be mitigated in the system. 

Moreover, in Table~\ref{table_fairness} we report the Jain fairness measure and the average \gls{ue} throughput varying the \gls{gnb} density, which is considered equal for all the operators in the area. 
We recall that, even if the densities of \glspl{ue} and \glspl{gnb} are fixed and equal for all the operators, the precise number of nodes deployed is random and follows two independent \glspl{ppp}.
As the table reports, with the dynamic use of the spectrum, the average throughput is slightly bigger than the baseline, but more importantly, the fairness increases, which means that resources are better assigned among all the \glspl{ue}.
Other spectrum sharing techniques (i.e., the ones studied in~\cite{rebato16,boccardi16,rebato_tccn}) can further improve the average throughput and the spectral efficiency, although drastically reducing the fairness among the \glspl{ue}.
Moreover, such schemes would require accurate coordination, which may be costly in dense networks.

\begin{table}[]
\centering
\caption{Comparison of the Jain fairness measure and the average \gls{ue} throughput $\bar{\eta}$ [Gbps] varying the \gls{gnb} density per operator.}
\centering \footnotesize
\renewcommand{\arraystretch}{0.98}
\begin{tabular}{lr|c|c|c}
\toprule
&               & 50 gNBs/km$^2$ & 75 gNBs/km$^2$ & 100 gNBs/km$^2$ \\ \midrule
\multirow{2}{*}{\colorbox{rgb:red!20,0.016;green!20,0.447;blue!20,0.694}{baseline} }& $\mathcal{J}$ & 0.6798         & 0.7384         & 0.7584          \\ \cline{2-5} 
                          & $\bar{\eta}$  & 0.2333         & 0.2422         & 0.2475          \\ \hline
\multirow{2}{*}{\colorbox{yellow!20}{dynamic}}  &  $\mathcal{J}$    & 0.8406         & 0.8719         & 0.8834          \\ \cline{2-5} 
                          & $\bar{\eta}$  & 0.2359         & 0.2510         & 0.2527   \\ \bottomrule  
\end{tabular}
\label{table_fairness}
\end{table}

\vspace{-0.35cm}
\section{Conclusion and future works}
\label{conclusion_and_future_works}
In this paper, we suggest the use of an \gls{lsa} or a third-party spectrum broker approach to dynamically share the total system band among operators in 5G mmWave cellular networks.
In the context of the Italian regulator, which 	in turn follows the European directives, we suggest the use of a dynamic allocation of resources among operators.
It appears useful to use a dynamic sharing approach for these frequencies in a way to better manage the spectrum and improve \gls{ue} fairness.
Our preliminary results show that dynamically sharing the spectrum according to the number of \glspl{ue} associated to each operator results in better fairness among the \glspl{ue} and also in the improvement of both spectral efficiency and user \gls{qos}.

As a future extension of this work, we will further study spectrum sharing approaches in line with the specifications of the regulators.
An improvement of the preliminary analysis is required, focusing also on other specific outcomes of the auctions between operators.
Moreover, as previously mentioned, a further economic study is needed in order to optimize the use of spectrum considering also a cost model for the operators that use the shared bands.

\vspace{-0.33cm}
\bibliographystyle{IEEEtran}
\bibliography{biblio}

\end{document}

%% file: acronyms.tex
\newacronym{laa}{LAA}{Licensed Assisted Access}
\newacronym{su}{SU}{Secondary User}
\newacronym{pu}{PU}{Primary User}
\newacronym{fss}{FSS}{Fixed-Satellite Service}
\newacronym{upa}{UPA}{Uniform Planar Array}
\newacronym{qos}{QoS}{Quality of Service}
\newacronym{lsa}{LSA}{Licensed Spectrum Access}
\newacronym{3gpp}{3GPP}{3rd Generation Partnership Project}
\newacronym{adc}{ADC}{Analog to Digital Converter}
\newacronym{5g}{5G}{5th generation}
\newacronym{aimd}{AIMD}{Additive Increase Multiplicative Decrease}
\newacronym{am}{AM}{Acknowledged Mode}
\newacronym{amc}{AMC}{Adaptive Modulation and Coding}
\newacronym{aqm}{AQM}{Active Queue Management}
\newacronym{awgn}{AGWN}{Additive White Gaussian Noise}
\newacronym{balia}{BALIA}{Balanced Link Adaptation}
\newacronym{bdp}{BDP}{Bandwidth-Delay Product}
\newacronym{bf}{BF}{Beamforming}
\newacronym{cc}{CC}{Congestion Control}
\newacronym{cdf}{CDF}{Cumulative Distribution Function}
\newacronym{cn}{CN}{Core Network}
\newacronym{cqi}{CQI}{Channel Quality Information}
\newacronym{cp}{CP}{Control Plane}
\newacronym{csirs}{CSI-RS}{Channel State Information - Reference Signal}
\newacronym{dc}{DC}{Dual Connectivity}
\newacronym{dce}{DCE}{Direct Code Execution}
\newacronym{dci}{DCI}{Downlink Control Information}
\newacronym{dl}{DL}{Downlink}
\newacronym{dmr}{DMR}{Deadline Miss Ratio}
\newacronym{dmrs}{DMRS}{DeModulation Reference Signal}
\newacronym{e2e}{E2E}{End-to-End}
\newacronym{ecn}{ECN}{Explicit Congestion Notification}
\newacronym{edf}{EDF}{Earliest Deadline First}
\newacronym{enb}{eNB}{evolved Node Base}
\newacronym{epc}{EPC}{Evolved Packet Core}
\newacronym{es}{ES}{Edge Server}
\newacronym{fdma}{FDMA}{Frequency Division Multiple Access}
\newacronym{fdd}{FDD}{Frequency Division Duplexing}
\newacronym[firstplural=Radio Access Technologies (RATs)]{rat}{RAT}{Radio Access Technology}
\newacronym{fs}{FS}{Fast Switching}
\newacronym{ftp}{FTP}{File Transfer Protocol}
\newacronym{gnb}{gNB}{Next Generation Node Base}
\newacronym{harq}{HARQ}{Hybrid Automatic Repeat reQuest}
\newacronym{hetnet}{HetNet}{Heterogeneous Network}
\newacronym{hh}{HH}{Hard Handover}
\newacronym{hol}{HOL}{Head-of-Line}
\newacronym{ia}{IA}{Initial Access}
\newacronym{imt}{IMT}{International Mobile Telecommunication}
\newacronym{iot}{IoT}{Internet of Things}
\newacronym{los}{LOS}{Line of Sight}
\newacronym{lte}{LTE}{Long Term Evolution}
\newacronym{m2m}{M2M}{Machine to Machine}
\newacronym{mac}{MAC}{Medium Access Control}
\newacronym{mc}{MC}{Multi-Connectivity}
\newacronym{mcs}{MCS}{Modulation and Coding Scheme}
\newacronym{mec}{MEC}{Mobile Edge Cloud}
\newacronym{mi}{MI}{Mutual Information}
\newacronym{mimo}{MIMO}{Multiple Input, Multiple Output}
\newacronym{mmwave}{mmWave}{millimeter wave}
\newacronym{mptcp}{MPTCP}{Multipath TCP}
\newacronym{mr}{MR}{Maximum Rate}
\newacronym{mss}{MSS}{Maximum Segment Size}
\newacronym{mtd}{MTD}{Machine-Type Device}
\newacronym{mtu}{MTU}{Maximum Transmission Unit}
\newacronym{nfv}{NFV}{Network Function Virtualization}
\newacronym{nlos}{NLOS}{Non Line of Sight}
\newacronym{nr}{NR}{New Radio}
\newacronym{ofdm}{OFDM}{Orthogonal Frequency Division Multiplexing}
\newacronym{pdcch}{PDCCH}{Physical Downlonk Control Channel}
\newacronym{pdcp}{PDCP}{Packet Data Convergence Protocol}
\newacronym{pdsch}{PDSCH}{Physical Downlink Shared Channel}
\newacronym{pdu}{PDU}{Packet Data Unit}
\newacronym{pf}{PF}{Proportional Fair}
\newacronym{pgw}{PGW}{Packet Gateway}
\newacronym{phy}{PHY}{Physical}
\newacronym{pbch}{PBCH}{Physical Broadcast Channel}
\newacronym[plural=\gls{mme}s,firstplural=Mobility Management Entities (MMEs)]{mme}{MME}{Mobility Management Entity}
\newacronym{prb}{PRB}{Physical Resource Block}
\newacronym{pss}{PSS}{Primary Synchronization Signal}
\newacronym{pucch}{PUCCH}{Physical Uplink Control Channel}
\newacronym{pusch}{PUSCH}{Physical Uplink Shared Channel}
\newacronym{rach}{RACH}{Random Access Channel}
\newacronym{ran}{RAN}{Radio Access Network}
\newacronym{red}{RED}{Random Early Detection}
\newacronym{rf}{RF}{Radio Frequency}
\newacronym{rlc}{RLC}{Radio Link Control}
\newacronym{rlf}{RLF}{Radio Link Failure}
\newacronym{rrc}{RRC}{Radio Resource Control}
\newacronym{rrm}{RRM}{Radio Resource Management}
\newacronym{rr}{RR}{Round Robin}
\newacronym{rs}{RS}{Remote Server}
\newacronym{rsrp}{RSRP}{Reference Signal Received Power}
\newacronym{rss}{RSS}{Received Signal Strength}
\newacronym{rtt}{RTT}{Round Trip Time}
\newacronym{rw}{RW}{Receive Window}
\newacronym{rx}{RX}{Receiver}
\newacronym{sa}{SA}{standalone}
\newacronym{sack}{SACK}{Selective Acknowledgment}
\newacronym{sap}{SAP}{Service Access Point}
\newacronym{sch}{SCH}{Secondary Cell Handover}
\newacronym{scoot}{SCOOT}{Split Cycle Offset Optimization Technique}
\newacronym{sdma}{SDMA}{Spatial Division Multiple Access}
\newacronym{sinr}{SINR}{Signal to Interference plus Noise Ratio}
\newacronym{sm}{SM}{Saturation Mode}
\newacronym{snr}{SNR}{Signal to Noise Ratio}
\newacronym{son}{SON}{Self-Organizing Network}
\newacronym{ss}{SS}{Synchronization Signal}
\newacronym{srs}{SRS}{Sounding Reference Signal}
\newacronym{sss}{SSS}{Secondary Synchronization Signal}
\newacronym{tb}{TB}{Transport Block}
\newacronym{tcp}{TCP}{Transmission Control Protocol}
\newacronym{tdd}{TDD}{Time Division Duplexing}
\newacronym{tdma}{TDMA}{Time Division Multiple Access}
\newacronym{tfl}{TfL}{Transport for London}
\newacronym{tm}{TM}{Transparent Mode}
\newacronym{trp}{TRP}{Transmitter Receiver Pair}
\newacronym{tti}{TTI}{Transmission Time Interval}
\newacronym{ttt}{TTT}{Time-to-Trigger}
\newacronym{tx}{TX}{Transmitter}
\newacronym{ue}{UE}{User Equipment}
\newacronym{ul}{UL}{Uplink}
\newacronym{uml}{UML}{Unified Modeling Language}
\newacronym{um}{UM}{Unacknowledged Mode}
\newacronym{utc}{UTC}{Urban Traffic Control}
\newacronym{vm}{VM}{Virtual Machine}
\newacronym{rsrq}{RSRQ}{Reference Signal Received Quality}
\newacronym{rssi}{RSSI}{Received Signal Strength Indicator}
\newacronym{crs}{CRS}{Cell Reference Signal}
\newacronym{nsa}{NSA}{Non Stand Alone}
\newacronym{mrdc}{MR-DC}{Multi \gls{rat} \gls{dc}}
\newacronym{endc}{EN-DC}{E-UTRAN-\gls{nr} \gls{dc}}
\newacronym{5gc}{5GC}{5G Core}
\newacronym{si}{SI}{Study Item}
\newacronym{iab}{IAB}{Integrated Access and Backhaul}
\newacronym{wf}{WF}{Wired-first}
\newacronym{hqf}{HQF}{Highest-quality-first}
\newacronym{pa}{PA}{Position-aware}
\newacronym{mlr}{MLR}{Maximum-local-rate}
\newacronym{wbf}{WBF}{Wired Bias Function}
\newacronym{mib}{MIB}{Master Information Block}
\newacronym{sib}{SIB}{Secondary Information Block}
\newacronym{kpi}{KPI}{Key Performance Indicator}
\newacronym[plural=PPPs,firstplural=Poisson Point Processes (PPPs)]{ppp}{PPP}{Poisson Point Process}
\newacronym{vr}{VR}{Virtual Reality}
\newacronym{scm}{SCM}{Spatial Channel Model}

%% file: AGCOM_arxiv_version.bbl
\begin{thebibliography}{10}
\providecommand{\url}[1]{#1}
\csname url@samestyle\endcsname
\providecommand{\newblock}{\relax}
\providecommand{\bibinfo}[2]{#2}
\providecommand{\BIBentrySTDinterwordspacing}{\spaceskip=0pt\relax}
\providecommand{\BIBentryALTinterwordstretchfactor}{4}
\providecommand{\BIBentryALTinterwordspacing}{\spaceskip=\fontdimen2\font plus
\BIBentryALTinterwordstretchfactor\fontdimen3\font minus
  \fontdimen4\font\relax}
\providecommand{\BIBforeignlanguage}[2]{{%
\expandafter\ifx\csname l@#1\endcsname\relax
\typeout{** WARNING: IEEEtran.bst: No hyphenation pattern has been}%
\typeout{** loaded for the language `#1'. Using the pattern for}%
\typeout{** the default language instead.}%
\else
\language=\csname l@#1\endcsname
\fi
#2}}
\providecommand{\BIBdecl}{\relax}
\BIBdecl

\bibitem{cisco2017}
Cisco, ``{Cisco Visual Networking Index: Global Mobile Data Traffic Forecast
  Update, 2016–2021},'' \emph{White Paper}, March 2017.

\bibitem{rangan2017potentials}
S.~Rangan, T.~S. Rappaport, and E.~Erkip, ``{Millimeter-Wave Cellular Wireless
  Networks: Potentials and Challenges},'' \emph{Proceedings of the IEEE}, vol.
  102, no.~3, pp. 366--385, March 2014.

\bibitem{rspg18}
{Radio spectrum policy group}, ``{Strategic spectrum roadmap towards 5G for
  Europe, RSPG Second Opinion on 5G networks, RSPG18-005 FINAL},'' Tech. Rep.,
  2018.

\bibitem{guidolin}
F.~Guidolin, M.~Nekovee, L.~Badia, and M.~Zorzi, ``{A study on the coexistence
  of fixed satellite service and cellular networks in a mmWave scenario},'' in
  \emph{IEEE International Conference on Communications (ICC)}, June 2015, pp.
  2444--2449.

\bibitem{massaro}
M.~Massaro, ``{Next generation of radio spectrum management: Licensed shared
  access for 5G},'' \emph{Telecommunications Policy}, vol.~41, no.~5, pp. 422
  -- 433, 2017.

\bibitem{future_spectrum}
R.~Umar, A.~U.~H. Sheikh, M.~Deriche, M.~Shoaib, and M.~Hadi, ``{Multi-operator
  spectrum sharing in next generation wireless communications networks: A short
  review and roadmap to future},'' in \emph{International Symposium on Wireless
  Systems and Networks (ISWSN)}, Nov 2017.

\bibitem{Bhattarai}
S.~Bhattarai, J.~M.~J. Park, B.~Gao, K.~Bian, and W.~Lehr, ``An {O}verview of
  {D}ynamic {S}pectrum {S}haring: {O}ngoing {I}nitiatives, {C}hallenges, and a
  {R}oadmap for {F}uture {R}esearch,'' \emph{IEEE Transactions on Cognitive
  Communications and Networking}, vol.~2, no.~2, pp. 110--128, June 2016.

\bibitem{survey}
R.~H. Tehrani, S.~Vahid, D.~Triantafyllopoulou, H.~Lee, and K.~Moessner,
  ``{Licensed Spectrum Sharing Schemes for Mobile Operators: A Survey and
  Outlook},'' \emph{IEEE Communications Surveys and Tutorials}, vol.~18, no.~4,
  pp. 2591--2623, Fourth quarter 2016.

\bibitem{rebato16}
M.~Rebato, M.~Mezzavilla, S.~Rangan, and M.~Zorzi, ``Resource {S}haring in 5{G}
  mm{W}ave {C}ellular {N}etworks,'' in \emph{IEEE Conference on Computer
  Communications Workshops (INFOCOM WKSHPS)}, April 2016, pp. 271--276.

\bibitem{boccardi16}
F.~Boccardi, H.~Shokri-Ghadikolaei, G.~Fodor, E.~Erkip, C.~Fischione,
  M.~Kountouris, P.~Popovski, and M.~Zorzi, ``Spectrum {P}ooling in {M}m{W}ave
  {N}etworks: {O}pportunities, {C}hallenges, and {E}nablers,'' \emph{IEEE
  Communications Magazine}, vol.~54, no.~11, pp. 33--39, November 2016.

\bibitem{rebato_tccn}
M.~Rebato, F.~Boccardi, M.~Mezzavilla, S.~Rangan, and M.~Zorzi, ``{Hybrid
  Spectrum Sharing in mmWave Cellular Networks},'' \emph{IEEE Transactions on
  Cognitive Communications and Networking}, vol.~3, no.~2, pp. 155--168, June
  2017.

\bibitem{dynamic_hspsh}
A.~Lertsinsrubtavee, N.~Malouch, and S.~Fdida, ``Hybrid spectrum sharing
  through adaptive spectrum handoff for cognitive radio networks,'' in
  \emph{IFIP Networking Conference}, June 2014, pp. 1--9.

\bibitem{merwaday}
A.~Merwaday, M.~Yuksel, T.~Quint, I.~Güvenç, W.~Saad, and N.~Kapucu,
  ``Incentivizing spectrum sharing via subsidy regulations for future wireless
  networks,'' \emph{Computer Networks}, vol. 135, pp. 132--146, 2018.

\bibitem{luo}
Y.~Luo, L.~Gao, and J.~Huang, ``{Spectrum broker by geo-location database},''
  in \emph{IEEE Global Communications Conference (GLOBECOM)}, Dec 2012, pp.
  5427--5432.

\bibitem{agcom_doc}
{Allegato B alla delibera n. 89/18/CONS}, ``{C}onsultazione pubblica sulle
  procedure per l’assegnazione e le regole per l’utilizzo delle frequenze
  disponibili nelle bande 694-790 {MH}z, 3600-3800 {MH}z e 26.5-27.5 {GH}z per
  sistemi terrestri di comunicazioni elettroniche al fine di favorire la
  transizione verso la tecnologia 5{G}, ai sensi della legge 27 dicembre 2017,
  n. 205,'' Tech. Rep., 2018.

\bibitem{akdeniz14}
M.~Akdeniz, Y.~Liu, M.~Samimi, S.~Sun, S.~Rangan, T.~Rappaport, and E.~Erkip,
  ``Millimeter {W}ave {C}hannel {M}odeling and {C}ellular {C}apacity
  {E}valuation,'' \emph{IEEE Journal on Selected Areas in Communications},
  vol.~32, no.~6, pp. 1164--1179, June 2014.

\bibitem{rebato18}
M.~Rebato, L.~Resteghini, C.~Mazzucco, and M.~Zorzi, ``Study of realistic
  antenna patterns in {5G} mmwave cellular scenarios,'' in \emph{IEEE
  International Conference on Communications (ICC)}, Kansas City, USA, May
  2018.

\bibitem{jain}
R.~Jain, D.~M. Chiu, and W.~Hawe, ``{A Quantitative Measure of Fairness and
  Discrimination for Resource Allocation in Shared Computer Systems},''
  \emph{DEC Research Report TR-301}, 1984.

\end{thebibliography}
